\documentclass{article}

\usepackage{PRIMEarxiv}

\usepackage[utf8]{inputenc} 
\usepackage[T1]{fontenc}    
\usepackage{hyperref}       
\usepackage{url}            
\usepackage{booktabs}       
\usepackage{amsfonts}       
\usepackage{nicefrac}       
\usepackage{microtype}      
\usepackage{lipsum}
\usepackage{subfigure}
\usepackage{graphicx}
\usepackage[linesnumbered,ruled,vlined]{algorithm2e}
\usepackage{fancyhdr}       
\usepackage{graphicx}       
\graphicspath{{media/}}     

\title{Learning state machines via efficient hashing of future traces
}

\author{
  Robert Baumgartner, Sicco Verwer \\
  Intelligent Systems \\
  Delft University of Technology \\
  Delft, The Netherlands\\
  \texttt{\{r.baumgartner-1, s.e.verwer\}@tudelft.nl} \\
}

\begin{document}
\maketitle

\begin{abstract}
State machines are popular models to model and visualize discrete systems such as software systems, and to represent regular grammars. 
Most algorithms that passively learn state machines from data 
assume all the data to be 
available from the beginning and they load this data into memory. This makes it hard to apply them to continuously streaming data and results in large memory requirements when dealing with large datasets. 
In this paper we propose a method to learn state machines from data streams using the count-min-sketch data structure to reduce memory requirements. We apply state merging using 
the well-known red-blue-framework to reduce the search space. We implemented our approach in an established framework for learning state machines, and evaluated it on a well know dataset to provide experimental data, showing the effectiveness of our approach with respect to quality of the results and run-time.
\end{abstract}


\section{Introduction}
\label{sec:intro}
State machines are a well known means to model discrete systems and regular grammars. When learned from data, they provide a means to describe the underlying dynamics, as well as a method to visualize the behavior \cite{chris_interpreting}. State machines can be learned in different manners from data. One of the most well known ways to identify state machines is the evidence-driven-state-merging algorithm, a method using statistical evidence to find similarities in states of a state machine based on heuristics. We assume familiarity with standard algorithms for learning state machines from data. We refer the reader to~\cite{higuera} for an introduction.
A drawback of most state machine learning algorithms, including evidence-driven-state-merging, is the fact that they require all input data at once in one pass, thus they cannot be learned in an adaptive manner. This also leads to a large memory footprint for large inputs, since data has to be stored all at once. Some of the few works tackling these two issues are from \cite{balle_2012, balle2014,schmidt2014online}. In this paper we propose an alternative approach to overcome these two limitations, and introduce a heuristic that makes use of count-min-sketches to efficiently learn state machines from data-streams in an adaptive manner. 

In each state, a count-min-sketch stores counts of hashes of observed futures. These futures are either subsequences starting in that state or sliding windows. Storing only counts allows us to process data streams, as we can forget most of the future sequences that are possible after reaching a state. The count-min-sketch should of course provide a sufficiently good estimate of a states' future behavior. We therefore only allow merges between states with sufficient counts, determined by a threshold. To reduce memory, we only store the red core, the blue fringe, and the first layer of white states. As a consequence, the state-merging methods are much more efficient as every merge only induces a handful of additional ones due to the determinization/folding routines. The behavior in all other states is estimated by the sketches. Although this greatly reduces the number of states in memory, we show in experiments on the PAutomaC dataset~\cite{verwer_pautomac} that this provides good estimates.
All of our source code will be released open source as part of the FlexFringe tool~\cite{flexfringe}. 


\section{Related Work}\label{section:related_work}

In the literature, several approaches exist for learning state machines. Active learning learns a model via actively asking queries~\cite{queries, vaandrager2017model}, where the learner can actively make queries to the system to extract information and pose a hypothesis. A drawback of this method however is that it assumes the presence of an oracle that processes and answers the queries asked. 
An efficient approach to learn from traces is~\cite{SATsolver}, where they reformulated the problem and used SAT-solvers to learn deterministic finite automata (DFA). The method is shown to solve the problem exactly, however, it becomes inefficient on larger datasets. 
Another approach to learn state machines from input traces is by means of state merging. In this case a prefix tree, the so called Augmented Prefix Tree Acceptor (APTA), is built, which describes the set of input traces exactly. In line with Occam's razor the goal of state merging then is to minimize the APTA while still representing the set of input traces. The algorithm achieves this via finding pairs of states that show similar behavior and then merging them. Although shown to be NP-hard~\cite{complexity}, much research uses a state merging method, see e.g.,~\cite{learning_grammars}. 

A popular approach is the evidence-driven-state-merging algorithm (EDSM) \cite{lang_1998}. The classical Alergia algorithm \cite{alergia_1994} is a version of state merging that uses statistical tests to compare and merge. The k-tails algorithm performs merges but requires identical futures up to a given depth k \cite{ktails}. Updated search procedures improve the quality of state machines as well as improve run-time \cite{search1, search2, search3}. Other types of state machines can be learned via specialized algorithms, such as timed automata via the RTI algorithm ~\cite{verwer_rti} or a likelihood-ratio test~\cite{likelihood}, or learning extended finite state machines~\cite{walkinshaw_2016} or guarded finite state machines~\cite{gkplus}. Vodenčarević et al. proposed an algorithm that learns a hybrid state machine model by taking different aspects of a system into account during the learning process~\cite{vodencarevic_2011}.

Despite the vast amount of research done on learning state machines, most of the state merging procedures assume that the complete data is available at the start of the program. Little work can be found on learning state machines in a streamed fashion. In~\cite{schmidt2014online} a method is presented that uses frequent pattern data stream mining techniques to make a streaming state machine learner. \cite{balle_2012} learns state machines using modified Space-Saving sketches, and prove properties such as convergence and memory consumption. This work was extended with a parameter search strategy in~\cite{balle2014}. In ~\cite{schouten2018} a streamed merging method method was implemented in the Apache framework, also based on count-min-sketches to hash futures. 
Our approach is largely inspired by these two works. 

\section{Background}\label{section:background}


\subsection{Probabilistic deterministic finite automata}

A PDFA is a tuple $\mathcal{A} = \{\Sigma, Q, q_0, \delta, S, F\}$, where $\Sigma$ is a finite alphabet, $Q$ is a finite set of states, $q_0 \in Q$ is a unique starting state, $\delta : Q \times \Sigma \to Q \cup \{0\}$ is the transition function, $S : Q \times \Sigma \to \left[0,1\right]$ is the symbol probability function, and $F : Q \to \left[0,1\right]$ is the final probability function, such that $F(q) + \sum_{a \in \Sigma} S(q,a) = 1$ for all $q \in Q$.
Given a sequence of symbols $s = a_1, \ldots, a_n$, a PDFA can be used to compute the probability for the given sequence: $\mathcal{A}(s) = S(q_0,a_1) \cdot S(q_1, a_2) \cdot \ldots \cdot S(q_{n-1}, a_n) \cdot F(q_n)$, where $\delta(q_i, a_{i+1}) = q_{i+1}$ for all $0 \leq i \leq n-1$. 
A PDFA is called probabilistic because it assigns probabilities based on the symbol $S$ and final $F$ probability functions. It is called deterministic because the transition function $\delta$ (and hence its structure) is deterministic. A PDFA computes a probability distribution over $\Sigma^*$, i.e., $\sum_{s \in \Sigma^*} \mathcal{A}(s) = 1$. 


\subsection{Learning State Machine Models}

We focus on learning state machines via state merging~\cite{lang_1998}. Classical state merging starts by constructing a tree representing the input data, also called the APTA. The APTA is a tree accepting the input traces exactly. The goal of state merging then is to minimize this APTA iteratively while still representing the data. This is analogous to the generalization process in other machine learning algorithms. Minimization is done via comparing pairs of states on behavior and merging states with similar futures. 
Since in each step multiple merges can be possible, the heuristic computes a score. The highest scoring possible merge will be performed. In our work, we also employ the red-blue framework. The red-blue framework maintains a core of red states, with initially only the root state red. Non-red states that have a direct transition from a red state are called blue states, states that are neither red nor blue are called white. A candidate merge pair can then only be a pair of a red and a blue state. 


\section{Methodology}\label{section:methodology}


\subsection{The heuristic}\label{subsec:heuristic}

\begin{figure}[htbp]
    \centering
    \subfigure[Before store.]{\label{fig:cms_before}
      \includegraphics[width=0.4\textwidth]{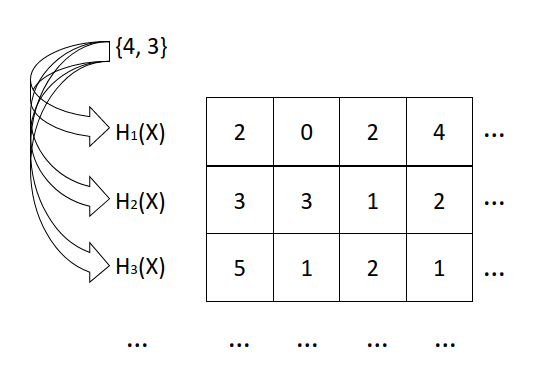}}
    \qquad
    \subfigure[After store.]{\label{fig:cms_after}
      \includegraphics[width=0.4\textwidth]{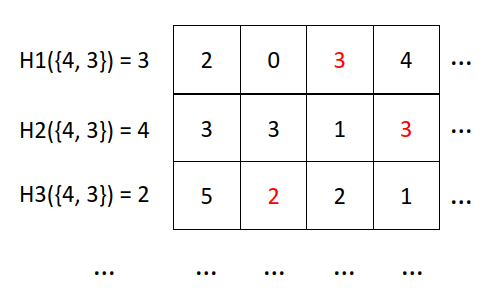}}
    \caption{A CMS before and after storing an element. The indices the hash functions hash the element onto are shown to the left.}
    \label{fig:csm_count}
\end{figure}

In order to efficiently store the future of states, our algorithm uses the 
count-min-sketch (CMS) data structure~\cite{count-min-sketch}. The main idea is to store the states' future with constant memory footprint per state, at the potential cost of approximation errors. It works via using a matrix, and the columns represent counts of hashed items. Since the hashing can collide, multiple rows are instantiated, each with its own associated hash function. Upon retrieving counts of an element $e$ off the CMS, $e$ is hashed again for each row, and the minimum count in the columns is considered the best approximation of the counts. Fig. ~\ref{fig:csm_count} shows the storage of a CMS. An element, in this case a tuple {4, 3}, is being stored and thrown into each row's hash function respectively. The increased counts are highlighted in red.

\begin{figure}[!ht]
    \centering
    \includegraphics[width=0.6\textwidth]{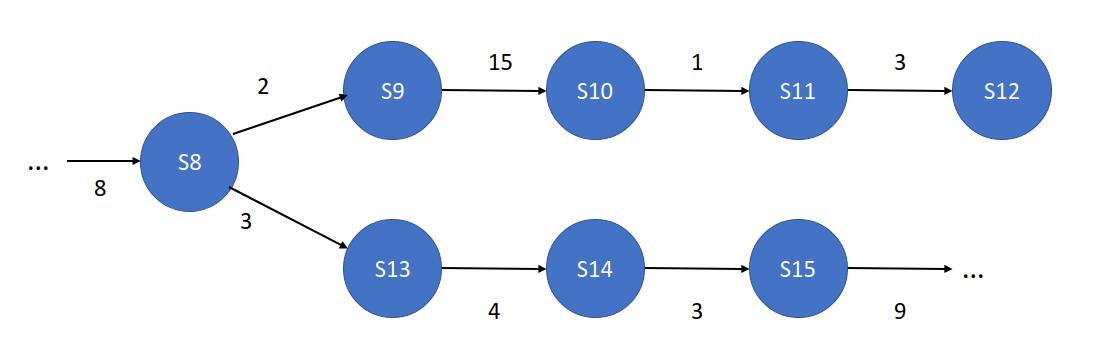}
    \caption{An excerpt of a prefix tree.}
    \label{fig:sketch_future}
\end{figure}

We use $n_{\mathit{futures}}$ CMS per state, one per future. If for example we take state $S8$ from Fig. ~\ref{fig:sketch_future} and set $n_{\mathit{futures}}$ to 3, the 3 sketches would look like the following: Sketch 1 would store a 2 and a 3, sketch 2 would store a 4 and a 15, sketch 3 would store a 3 and a 1. Additionally to the normal outgoing symbols, we save an extra dedicated bin in the sketch for terminating sequences. Whenever a sequence terminates in a state, we store a "sequence termination" on the last column of the sketch, and only terminations can be stored in that column.

Whenever we want to merge a state, all we have to do them is to compare the corresponding sketch pairs, i.e., sketch 1 of the red state with sketch 1 of the blue state, sketch 2 of the red state with sketch 2 of the blue state, and so on. We consider the rows of the CMS distributions, and perform the Hoeffding-bound similar to \cite{alergia_1994} (Eq.~\ref{eq:hoeffding}) for each corresponding pair of rows of the two sketches. In this equation, $\frac{x_i}{n_1}$ is the relative frequency of element $i$ in distribution $1$, and $\frac{y_i}{n_2}$ it's frequency in distribution $2$. $\alpha$ is a hyperparameter to be set. Apart from checking whether two states are can be merged, we also need a score function. In order to assign a score to a merge, we consider the rows of the CMS as vectors, and then compute the sum of the cosine-similarities of each corresponding vector (row) pair $\vec{v_i}^l$ and $\vec{v_i}^r$ (Eq.~\ref{eq:cosine_similarity}) in between two sketches, averaged over the number of rows each sketch has. Furthermore, when performing or undoing a merge on two states, we need to update the state information. To this end, we simply treat the sketches of the individual states as matrices, which we sum up on a merge.

\begin{equation}\label{eq:hoeffding}
    \left| \frac{x_i}{n_1} - \frac{y_i}{n_2} \right| < \sqrt{\frac{1}{2}log\left(\frac{2}{\alpha}\right)}\left(\frac{1}{\sqrt{n_1}}+\frac{1}{\sqrt{n_2}}\right),
\end{equation}

\begin{equation} \label{eq:cosine_similarity}
    cosine\_sim\left(\vec{v_i}^l, \vec{v_i}^r\right) = \frac{\vec{v_i}^l \cdot \vec{v_i}^r}{||\vec{v_i}^l|| ||\vec{v_i}^r||},
\end{equation}

\subsection{Streaming}\label{subsec:streaming}

In order to stream state machines we decided for an approach similar to the one in \cite{balle_2012}. We adopt the red-blue framework and include an evidence threshold $sink\_count$. Our streaming starts with the root node as a red node. Differently to the normal batch-wise approach we only create new states with direct transitions coming from red or blue states. We furthermore introduce a threshold for states. Every time an incoming sequence passes a state $S_i$, we increase a counter on that state by one. Only when the counter of a state passes a threshold can that state become a blue state and be considered for a merge. Merges happen only in batches. Once the batch size is reached, we perform merges until no more are possible, then we read the next batch. The streaming procedure is described in Alg.~\ref{alg:streaming}.

\begin{algorithm}[t]
  \label{alg:streaming}
  \caption{$Streaming$}
  \KwIn{Set of sequences $X$, batch size $b$, threshold $t$}
  \KwOut{A hypothesis (automaton) $H$}
  
  $H \leftarrow$ root node\;
  $c \leftarrow 0$\;
  \ForEach{$x_i \in X$}{
    $n \leftarrow$ root node\;
    \ForEach{$s_j \in x_i$}{
        // $s_j$ are the individual symbols of string $x_i$\;
        $n.size \leftarrow n.size + 1$\;
        \uIf{transition from $n$ with $s_j$ exists}{
            $n = \delta(n, s_j)$\;
        }
        \uElseIf{$n$ is red or $n$ is blue}{
            create new node satisfying $\delta(n, s_j)$\;
        }
        \uElseIf{$n.parent$ is red and $n.size == t$}{
            mark $n$ blue\;
        }
        \uElse{
            // this is a white state\;
            do nothing\;
        }
    }
    $c \leftarrow c + 1$\;
    \uIf{$c == b$}{
        \While{$merges\_possible(H)$}{
            $perform\_merges(H)$\;
        }
        $c\leftarrow 0$\;
    }
  }
  
\end{algorithm}

\section{Experiments and results}\label{section:experiments_results}

In order to evaluate our approach, we implemented our streaming and our heuristic as modules in Flexfringe \cite{flexfringe, flexfringeRepo}. In order to compare our approach, we compared it with a baseline approach, namely FlexFringe's implementation of the Alergia algorithm. The streaming procedure described in Alg.~\ref{alg:streaming} is generic enough to use any merge heuristic that supports a $consistency\_check$ and an $assign\_score$ procedure, and thus our implementation allows us to exchange the sketching and Alergia heuristic very easily. 
The dataset we used is the well known PAutomaC dataset \cite{verwer_pautomac}. We set the batchsize $b$ to $500$ and the threshold $t$ to $100$ to give the statistical tests enough evidence, and ran both the sketching with different future lengths $n_{\mathit{futures}}$, and the Alergia algorithm. In order to have a better comparison, we augmented the Alergia algorithm with the well known k-tails algorithm \cite{ktails}, from which we use the $k$ parameter. Since we only append to red and blue states, a value of $k$ greater than $1$ would be meaningless, hence we only used $k=0$ and $k=1$.

\begin{figure}
    \centering
    \subfigure[Alergia with varying $k$.]{\label{fig:alergia_ktails}
      \includegraphics[width=0.45\textwidth]{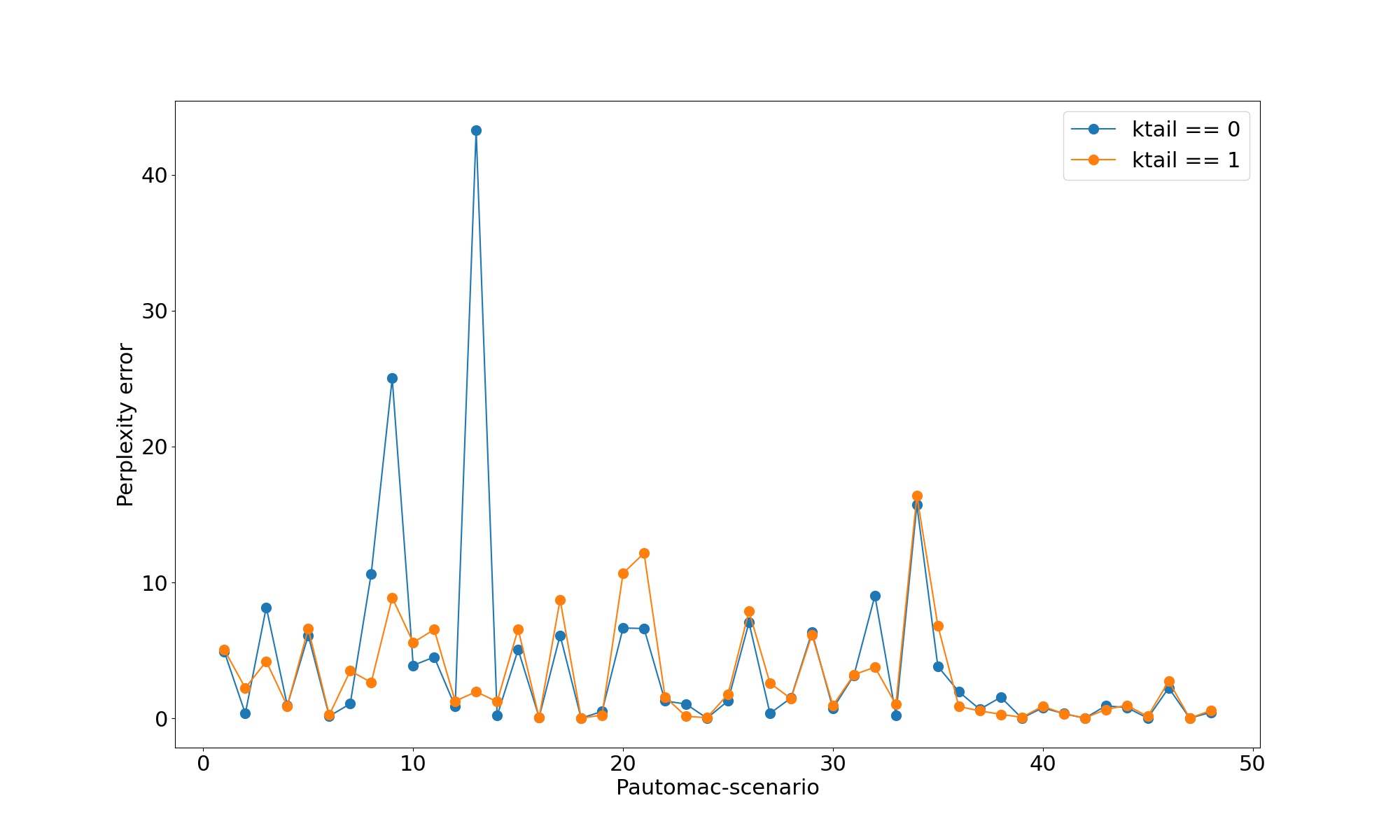}}
    \qquad
    \subfigure[Sketching heuristic with different $n_{\mathit{futures}}$ values.]{\label{fig:sketch_futures}
      \includegraphics[width=0.45\textwidth]{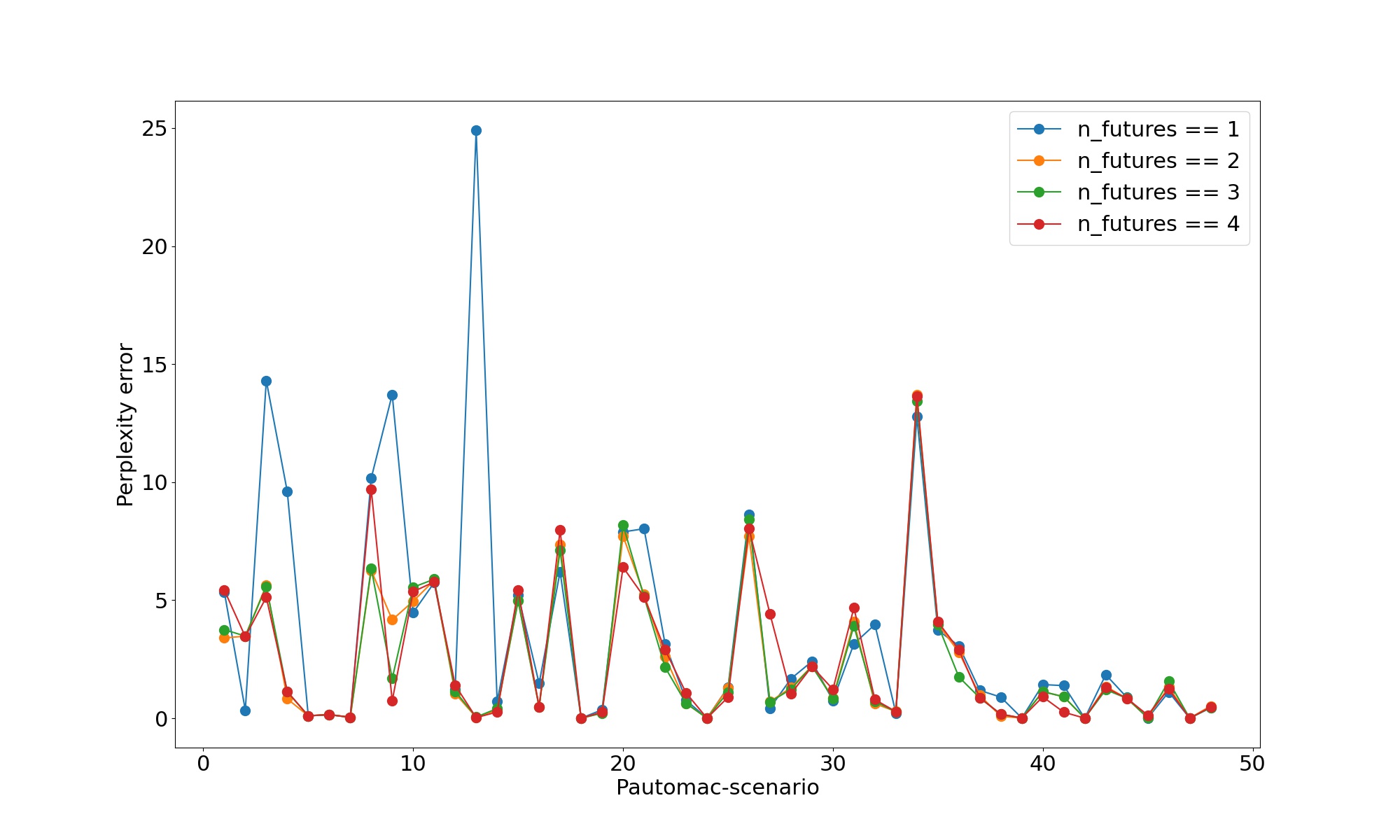}}
    \subfigure[Sketching heuristic with $n_{\mathit{futures}}=2$ vs. Alergia with $k=1$.]{\label{fig:sketch_vs_alergia}
      \includegraphics[width=0.45\textwidth]{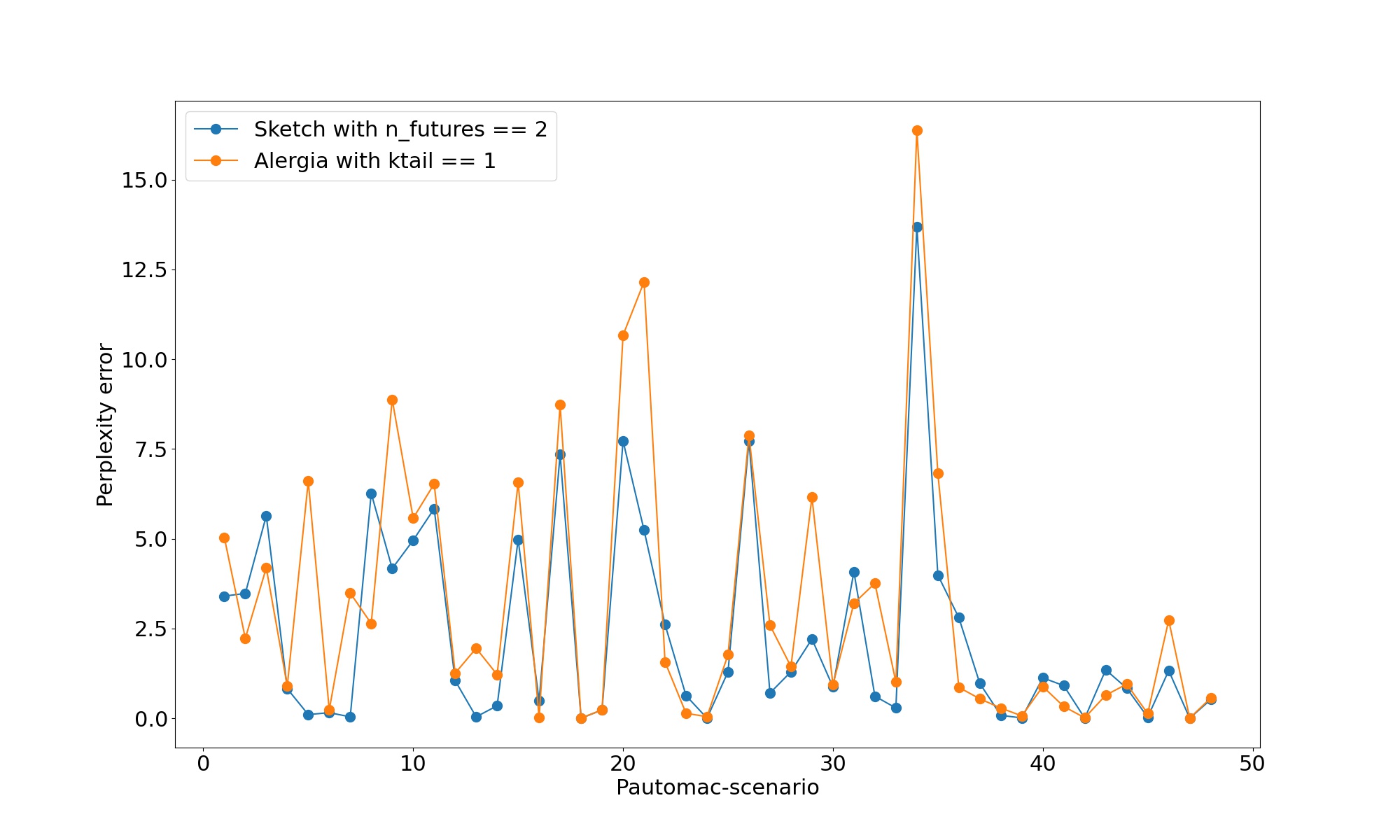}}
    \qquad
    \subfigure[Sketching heuristic with $n_{\mathit{futures}}=3$ in stream mode vs. Alergia with $k=2$ in batch mode.]{\label{fig:stream_vs_batch}
      \includegraphics[width=0.45\textwidth]{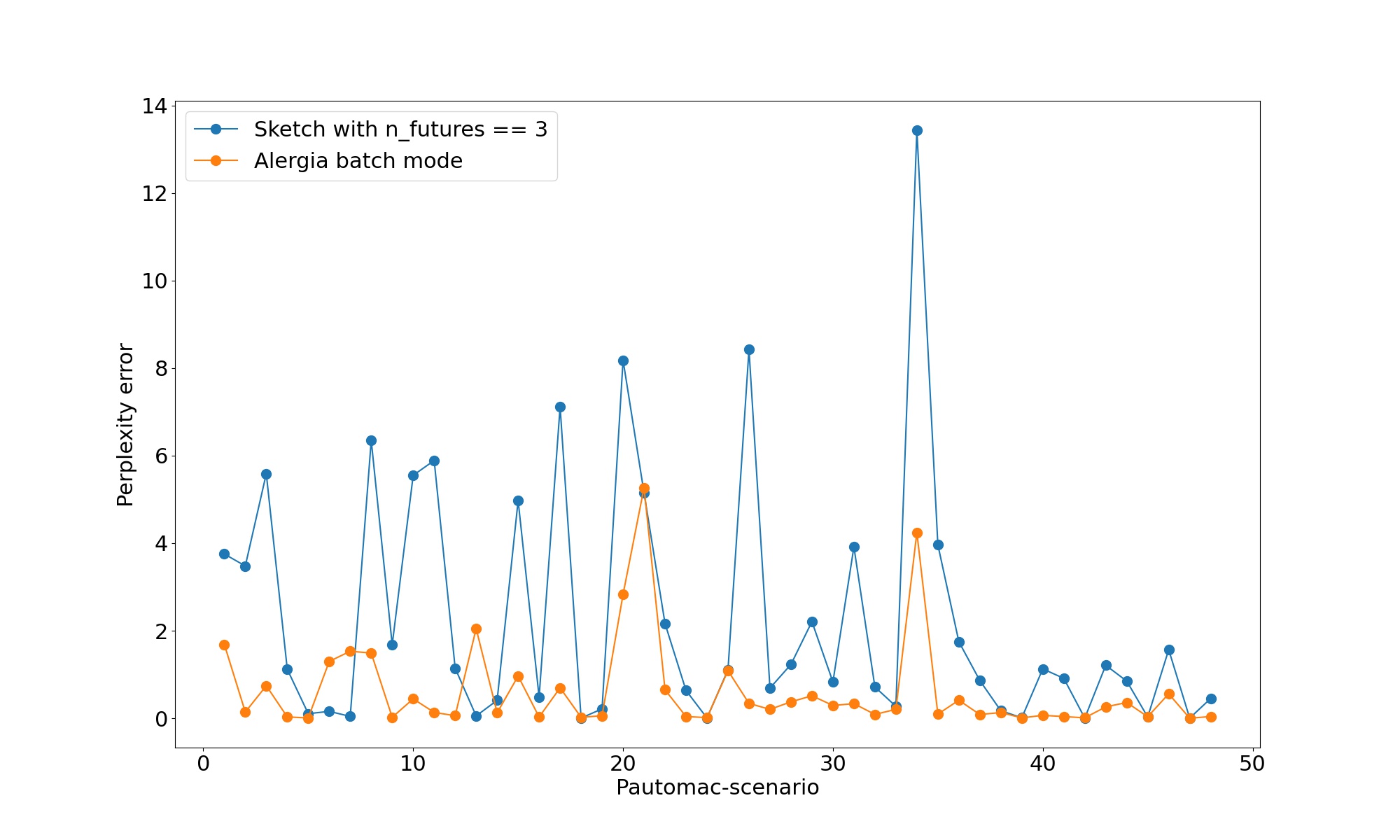}}
    \caption{Perplexity score errors on all experiments. The x-axis indicates the dataset's scenarios from 1 to 48}
\end{figure}

Our performance metric is the difference in perplexities of the original automaton and our learned automaton, as described in \cite{verwer_pautomac}.  Fig. ~\ref{fig:alergia_ktails} shows the streamed Alergia results with the two different values for $k$. It is clear that $k==1$ performs better.
In order to check how the sketching algorithm scales with respect to $n_{\mathit{futures}}$, we plot the perplexities in a similar fashion, where we only vary the $n_{\mathit{futures}}$ parameter this time. The plots are shown in Fig. ~\ref{fig:sketch_futures}. Increasing $n_{\mathit{futures}}$ from $1$ to $2$ has a large impact, similar to increasing $k$ from $0$ to $1$ in Alergia. Increasing the value to $3$ improves some problems minimally, however we noticed that from $n_{\mathit{futures}}=4$ on the quality of the resulting state machine starts to degrade. Also note that the sketching and Alergia have peaks at the same problems in their weaker settings, and both improve drastically on the same problems when giving them more information. 
In a next step we compare the sketching with Alergia in Fig. ~\ref{fig:sketch_vs_alergia}. We can see that the sketching performs better than Alergia on most problems, which is also evidenced by the average error of $2.34$ for the sketching in this setting and $3.14$ for Alergia. Increasing $n_{\mathit{futures}}$ to $3$ would result in an average error of $2.29$ in our experiment for the sketching, where it performed always equal or minimally better than the same with $n_{\mathit{futures}}=2$.

We also compared runtime, where we found almost identical times. While sketching took $36.8s$ in total on all problems with $n_{\mathit{futures}}=2$ on our machine (Ubuntu 20.4, Intel i7@2.60Ghz and 16GB RAM), Alergia took $36.0s$ with $k=1$. Increasing $n_{\mathit{futures}}$ to 3 increased the runtime to $38.8s$, the $k$ parameter had little effect on runtime from $0$ to $1$, with a runtime of $35.3s$ at $k=0$. Last but not least it is worth noting that while Alergia's performance increased with larger $k$, doing the same with the sketching approach actually resulted in worse results.
In order to compare the batch mode with the stream mode we ran Alergia in $k=2$ in batch mode and the sketching with $n_{\mathit{futures}}=3$ in stream mode. The comparison is depicted in Fig. ~\ref{fig:stream_vs_batch}. It can be seen that the streamed version compares worse than the batched version, which is also imminent from the average error of $0.63$ for the batch-mode. However, this comes at the cost of much larger memory and runtime consumption. While the streamed version ran for less than $40s$, the batch mode took $258s$, or more than $4$ minutes. Memory consumption wise the batched version is also much more costly, as can be seen in Table \ref{tab:memory_consumption}, where we compare memory footprint of batch mode Alergia and stream mode sketching on a few selected PAutomaC scenarios. To get the consumption we measured dynamic memory allocation via the Massif tool from the Valgrind toolkit.

\begin{table}
\centering
\begin{tabular}{ ||c||c|c|c|c|| }
 \hline
 Method & Scenario 8 & Scenario 9 & Scenario 20 & Scenario 21 \\
 \hline\hline
 Alergia batch & 1GB & 58.47MB & 957.7MB & 2.049GB \\ 
 Sketching stream & 95MB & 17.4MB & 35.53MB & 56.58MB \\
 \hline
\end{tabular}
\caption{Batch vs. stream mode memory consumption per maximum heap size for a few selected scenarios.}
\label{tab:memory_consumption}
\end {table}

\section{Discussion and limitations}\label{section:discussion}

At first we expect the gain in performance for Alergia with increasing $k$. In batch mode, up to a value of $k=2$ performance slightly increases, then it enters a plateau. Apparently, for the PAutomaC dataset, it is sufficient to know the next $3$ steps ahead. The decrease in the performance of our sketching with $n_{\mathit{futures}}>3$ can be possibly explained by the fact that the algorithm will prefer non-optimal merges, since the sketches "pool" the future subtrees together for the score computation.  

The advantage of our sketching approach lies in the fact that we can look ahead further than other heuristics like Alergia while still discarding information, i.e., the states themselves. We also approximate infrequently used symbols, unlike ~\cite{balle_2012}, who use a modification of the Space-Saving-Algorithm to approximate the most frequent features only. A drawback of our sketching is though that with too large alphabet size our approximations can collide too much, leading to worse results.  
We are not sure yet why our method's results degrade with higher $k$ larger than $0$. We can only provide a hypothesis at this stage. In order to do ktails and with our way of streaming, pairs of white states with transitions coming out of blue states have to be compared. Those states can, since they are usually infrequent, possibly not provide much evidence, leading to incorrect results on our statistical tests.
Limitations of this work are mainly the nature of the dataset, which appears simple enough that not many lookaheads and simple methods are enough to get good results. But our results to demonstrate that the approach works. Another limitation is the size of the dataset, which are small for streaming algorithms. In the future, we aim to test our method on different data sets such as network traffic and software logs, which are typically too large to fit into memory. 

\section{Conclusion}\label{section:conclusion}

In this work we introduced a new approach for streaming state machines and showed first experimental results. We compared our sketching approach with a conventional method and pointed out advantages and disadvantages based on the results we got. We conclude that our method works as expected, and we expect the advantages to show better on larger datasets where deeper lookaheads into the future of a state are necessary to make good predictions. The streaming also enables the processing of really large data in a cost effective manner. A drawback of our method however is clearly that the sketches have a fixed size when starting the algorithm, and thus the number of potential symbols the traces may contain must be known or estimated before running the algorithm. When set too small, the results could get impaired by collisions from inside the CMS. 

\section{Ackknowledgements}

This work is supported by NWO TTW VIDI project 17541 - Learning state machines from infrequent software traces (LIMIT).

\bibliographystyle{unsrt}  
\bibliography{references}

\begin{thebibliography}{10}

\bibitem{chris_interpreting}
Christian Hammerschmidt, Sicco Verwer, Qin Lin, and Radu State.
\newblock Interpreting finite automata for sequential data.
\newblock 12 2016.

\bibitem{higuera}
Colin de~la Higuera.
\newblock {\em Grammatical Inference: Learning Automata and Grammars}.
\newblock Cambridge University Press, 2010.

\bibitem{balle_2012}
Borja Balle, Jorge Castro, and Ricard Gavaldà.
\newblock Bootstrapping and learning pdfa in data streams.
\newblock In Jeffrey Heinz, Colin Higuera, and Tim Oates, editors, {\em
  Proceedings of the Eleventh International Conference on Grammatical
  Inference}, volume~21 of {\em Proceedings of Machine Learning Research},
  pages 34--48, University of Maryland, College Park, MD, USA, 05--08 Sep 2012.
  PMLR.

\bibitem{balle2014}
Borja Balle, Jorge Castro, and Ricard Gavald{\`{a}}.
\newblock {Adaptively learning probabilistic deterministic automata from data
  streams}.
\newblock {\em Mach Learn}, 96:99--127, 2014.

\bibitem{schmidt2014online}
Jana Schmidt and Stefan Kramer.
\newblock Online induction of probabilistic real-time automata.
\newblock {\em Journal of Computer Science and Technology}, 29(3):345--360,
  2014.

\bibitem{verwer_pautomac}
Sicco Verwer, R{\'{e}}mi Eyraud, and Colin {De La Higuera}.
\newblock {PAutomaC: A probabilistic automata and hidden Markov models learning
  competition}.
\newblock {\em Machine Learning}, 96(1-2):129--154, oct 2014.

\bibitem{flexfringe}
Sicco Verwer and Christian~A. Hammerschmidt.
\newblock flexfringe: A passive automaton learning package.
\newblock In {\em 2017 IEEE International Conference on Software Maintenance
  and Evolution (ICSME)}, pages 638--642, 2017.

\bibitem{queries}
Dana Angluin.
\newblock Queries and concept learning.
\newblock {\em Machine Learning}, 2:319--342, 1988.

\bibitem{vaandrager2017model}
Frits Vaandrager.
\newblock Model learning.
\newblock {\em Communications of the ACM}, 60(2):86--95, 2017.

\bibitem{SATsolver}
Marijn J.~H. Heule and Sicco Verwer.
\newblock Exact dfa identification using sat solvers.
\newblock In Jos{\'e}~M. Sempere and Pedro Garc{\'i}a, editors, {\em
  Grammatical Inference: Theoretical Results and Applications}, pages 66--79,
  Berlin, Heidelberg, 2010. Springer Berlin Heidelberg.

\bibitem{complexity}
E~Mark Gold.
\newblock Complexity of automaton identification from given data.
\newblock {\em Information and Control}, 37(3):302--320, 1978.

\bibitem{learning_grammars}
Colin {de la Higuera}.
\newblock A bibliographical study of grammatical inference.
\newblock {\em Pattern Recognition}, 38(9):1332--1348, 2005.
\newblock Grammatical Inference.

\bibitem{lang_1998}
Kevin~J. Lang, Barak~A. Pearlmutter, and Rodney~A. Price.
\newblock {Results of the abbadingo one DFA learning competition and a new
  evidence-driven state merging algorithm}.
\newblock In {\em Lecture Notes in Computer Science (including subseries
  Lecture Notes in Artificial Intelligence and Lecture Notes in
  Bioinformatics)}, volume 1433, pages 1--12. Springer Verlag, 1998.

\bibitem{alergia_1994}
Rafael~C. Carrasco and Jose Oncina.
\newblock Learning stochastic regular grammars by means of a state merging
  method.
\newblock In Rafael~C. Carrasco and Jose Oncina, editors, {\em Grammatical
  Inference and Applications}, pages 139--152, Berlin, Heidelberg, 1994.
  Springer Berlin Heidelberg.

\bibitem{ktails}
A.~W. Biermann and J.~A. Feldman.
\newblock On the synthesis of finite-state machines from samples of their
  behavior.
\newblock {\em IEEE Transactions on Computers}, C-21(6):592--597, 1972.

\bibitem{search1}
A.L. Oliveira and J.P.M. Silva.
\newblock Efficient search techniques for the inference of minimum size finite
  automata.
\newblock In {\em Proceedings. String Processing and Information Retrieval: A
  South American Symposium (Cat. No.98EX207)}, pages 81--89, 1998.

\bibitem{search2}
John Abela, François Coste, and Sandro Spina.
\newblock Mutually compatible and incompatible merges for the search of the
  smallest consistent dfa.
\newblock volume 3264, pages 28--39, 10 2004.

\bibitem{search3}
Kevin Lang.
\newblock Faster algorithms for finding minimal consistent dfas.
\newblock 01 2000.

\bibitem{verwer_rti}
Sicco Verwer, Mathijs Weerdt, and Cees Witteveen.
\newblock Efficiently identifying deterministic real-time automata from labeled
  data.
\newblock {\em Machine Learning}, 86:295--333, 03 2012.

\bibitem{likelihood}
Sicco Verwer, M.~D. Weerdt, and Cees Witteveen.
\newblock A likelihood-ratio test for identifying probabilistic deterministic
  real-time automata from positive data.
\newblock In {\em ICGI}, 2010.

\bibitem{walkinshaw_2016}
Neil Walkinshaw, Ramsay Taylor, and John Derrick.
\newblock {Inferring extended finite state machine models from software
  executions}.
\newblock {\em Empirical Software Engineering}, 21(3):811--853, 2016.

\bibitem{gkplus}
Leonardo Mariani, Mauro Pezzè, and Mauro Santoro.
\newblock Gk-tail+ an efficient approach to learn software models.
\newblock {\em IEEE Transactions on Software Engineering}, 43(8):715--738,
  2017.

\bibitem{vodencarevic_2011}
Asmir Vodenčarević, Hans~Kleine Bürring, Oliver Niggemann, and Alexander
  Maier.
\newblock Identifying behavior models for process plants.
\newblock In {\em ETFA2011}, pages 1--8, 2011.

\bibitem{schouten2018}
Hans Schouten.
\newblock {Learning State Machines from data streams and an application in
  network-based threat detection}.
\newblock Technical report, 2018.

\bibitem{count-min-sketch}
Graham Cormode and S.~Muthukrishnan.
\newblock An improved data stream summary: the count-min sketch and its
  applications.
\newblock {\em Journal of Algorithms}, 55(1):58--75, 2005.

\bibitem{flexfringeRepo}
Sicco Verwer and Christopher Hammerschmidt.
\newblock Flexfringe.
\newblock \url{https://github.com/tudelft-cda-lab/FlexFringe}.

\end{thebibliography}

\end{document}